\documentclass[twocolumn,preprintnumbers,amsmath,amssymb,showpacs]{revtex4}
\begin{document}
\title{Baryons in Massive Gross-Neveu Models}
\preprint{FAU-TP3-05/3}
\author{Michael Thies}
\author{Konrad Urlichs\footnote{Electronic addresses:
thies@theorie3.physik.uni-erlangen.de,  kon- rad@theorie3.physik.uni-erlangen.de}}
\affiliation{Institut f\"ur Theoretische Physik III,
Universit\"at Erlangen-N\"urnberg, D-91058 Erlangen, Germany}
\date{\today}
\begin{abstract}
Baryons in the large $N$ limit of (1+1)-dimensional Gross-Neveu models with either discrete or
continuous chiral symmetry have long been known. We generalize their construction to the case
where the symmetry is explicitly broken by a bare mass term in the Lagrangian. In the discrete symmetry case, the
exact solution is found for arbitrary bare fermion mass, using the Hartree-Fock approach. 
It is mathematically closely related to polarons and bipolarons in conducting polymers. In the
continuous symmetry case, a derivative expansion allows us to rederive a formerly proposed Skyrme-type
model and to compute systematically corrections to the leading order description
based on an effective sine-Gordon theory.  
\end{abstract}
\maketitle
%################################################################################################
%                                                                                           SECTION 1
%################################################################################################
\section{Introduction}
The original Gross-Neveu models \cite{1} are asymptotically free, self-interacting fermionic field theories
in 1+1 dimensions  with Lagrangian
\begin{equation}
{\cal L} = \bar{\psi} \left({\rm i} \gamma^{\mu}\partial_{\mu}-m_0\right)\psi  + \frac{1}{2} g^2 \left[ (\bar{\psi}\psi)^2+
 \eta (\bar{\psi} {\rm i}\gamma^5 \psi)^2\right] .
\label{i1}
\end{equation}
We suppress the flavor indices ($i=1...N$) and shall only consider the 't Hooft limit $N\to \infty, Ng^2=$ const.
The model with discrete chiral symmetry
[$\eta=0$ in Eq.~(\ref{i1})] will be referred to as Gross-Neveu model  (GN),
the one with continuous chiral symmetry ($\eta=1$) as Nambu--Jona-Lasinio model in two dimensions (NJL$_2$)
throughout this paper. 
Apart from being of theoretical interest in their own right, these models have been useful as a theoretical
laboratory for new methods and algorithms and to describe quasi-one-dimensional condensed matter systems. 
Whereas the massless models ($m_0=0$) have been rather comprehensively studied by now 
(see \cite{2,3} for pertinent 
review articles and \cite{4,5,6} for a recent update on the GN model),
results about the massive versions with an explicit symmetry breaking term ($m_0\neq 0$) are scarce and somewhat
scattered through the literature. In view of the fact that in nature quarks are massive and a lot of effort is 
presently devoted to computing chiral corrections, we believe that it is both worthwhile and timely to study more
systematically  the massive
Gross-Neveu models. In order to motivate our specific work, let us briefly summarize the present status of this
field.
   
The gap equation which determines the vacuum and the dynamical fermion mass $m$ can easily be generalized to finite
bare quark mass $m_0$. Apart from $m$, a second physical, renormalization group invariant parameter appears,
$m_0/Ng^2$. It enters the renormalized vacuum energy and observables like the $\bar{q}q$ scattering amplitude or the
mass of the ``pion" in the NJL$_2$ model \cite{7}. The $\sigma$-meson is unbound and disappears from the
spectrum for any $m_0 \neq 0$,  the $\pi$-meson becomes massive in the familiar way 
(Gell-Mann, Oakes, Renner relation). If one assumes unbroken translational invariance, one can
also infer the phase diagram in the ($T,\mu$) plane \cite{8}.
However, since we now know that this assumption is not justified at $m_0=0$ \cite{6}, this analysis 
cannot be trusted and needs to be repeated.

As far as baryons in the massive Gross-Neveu models are concerned, we are only aware of two variational calculations
so far. Salcedo et al. \cite{9} have studied both the 't Hooft model (large $N$ QCD$_2$ \cite{9a}) and the NJL$_2$ model,
comparing a lattice Hartree-Fock calculation with a variational ansatz in which the 
chiral phase of the fermions serves as effective low energy field. In the limit of 
small but finite bare quark masses, the authors find a sine-Gordon theory, the two-dimensional analogue of the Skyrme model
where baryon
number is generated through a topologically non-trivial pion field configuration. This elegant approach yields the
correct dependence of the pion mass on the quark mass in a very simple manner and can readily be generalized
to finite temperature and chemical potential \cite{10,2}.
In the case of the 't Hooft model, the authors of Ref. \cite{9} compare the soliton approach with the full
Hartree-Fock calculation and notice that it works very well, even away from the chiral limit. Unfortunately, such a
comparison was not done for the NJL$_2$ model, so that the accuracy of the sine-Gordon approach 
cannot be assessed in the case at hand.

In the non-chiral GN model, Feinberg and Zee \cite{11} have performed a variational calculation,
based on the scalar potential of the standard ($m_0=0$) kink-antikink baryon \cite{12}. They compute the energy of the baryon
and discuss the limiting cases of small and large bare quark masses. They conclude that their ansatz does not satisfy the
saddle point equation. Claims made in this paper about the non-existence of static bags in the massive GN model have later
been retracted by the authors \cite{13}.

Since one does not know the accuracy of these variational calculations, they are insufficient if one wishes
to use the Gross-Neveu models as testing ground for new theoretical approaches. 
Moreover, as should be clear from the above short summary, the general situation concerning the massive
Gross-Neveu models is far from satisfactory. We therefore decided to reconsider this whole issue more
systematically. In the present work, we begin such a study with a fresh look at the baryon in both the massive GN and the
NJL$_2$ models. Obviously, this is a prerequisite for a future study of the full phase structure of these models.
In Sect. II, we will present the analytical solution for the baryon in the discrete chiral massive Gross-Neveu model
for arbitrary quark masses, obtained along lines very similar to Ref. \cite{11}. In Sect. III, we turn to the 
NJL$_2$ model and show that the Skyrme-type approach \cite{9} can be identified with the 
leading order term of a systematic derivative expansion. We will also compute the three following higher order
corrections  in  the derivative expansion in closed analytical form. It seems to us that the problem of
baryonic matter at finite temperature away from the chiral limit should be tractable owing to these new
insights, but we leave such an investigation for the future. Sect. IV contains a brief summary and our conclusions
as well as a comment on the quantitative relationship between baryons in the massive GN model on the one hand and 
polarons, bipolarons and excitons in conducting polymers on the other hand.
%################################################################################################
%                                                                                           SECTION 2
%################################################################################################
\section{Baryons in the massive GN model --- exact results}

Here, we basically repeat the variational calculation of Ref. \cite{11}, using the Hartree-Fock approach.
We simply write down a scalar trial potential and prove its self-consistency. Since our trial potential has
the same shape as the $m_0=0$ baryon with partial filling of the valence level (a kink-antikink potential
well \cite{12}), we can take over many results from the corresponding Hartree-Fock calculation in the chiral limit \cite{14}. We
refer the reader to this last paper for more details and will concentrate on the differences caused by the
bare quark mass.

The trial scalar potential reads
\begin{equation}
S(x) = m \left[ 1 + y \left( \tanh \xi_-  - \tanh \xi_+ \right)\right]
\label{e1}
\end{equation}
with
\begin{equation}
\xi_{\pm} = ymx \pm \frac{1}{2} {\rm arctanh}\, y ,
\label{e1a}
\end{equation}
where $m$ is the physical fermion mass in the vacuum, $y\in [0,1]$ the only variational parameter.
The potential is the same as in the $m_0=0$ case, except that we will get a different relation
between the parameter $y$ and the occupation fraction $\nu=n/N$ of the valence state.
We first evaluate the baryon mass relative to the vacuum.
The calculation of the sum over single particles is identical to that in Ref. \cite{14}.
Hence the contribution from the negative energy continuum states ($E(k)=\sqrt{k^2+m^2}\,$),
\begin{eqnarray}
\Delta E_{\rm cont} &=& \frac{2Nym}{\pi} + 2Nym \int_{-\Lambda/2}^{\Lambda/2} \frac{{\rm d}k}{2\pi} \frac{1}{E(k)}
\nonumber \\
& & + \frac{2Nm}{\pi} \sqrt{1-y^2} \arctan \frac{\sqrt{1-y^2}}{y}
\nonumber \\
& = & \frac{2Nym}{\pi}\left(1 +  \ln \frac{\Lambda}{m} \right.
\nonumber \\
& & \left. + \frac{\sqrt{1-y^2}}{y} \arctan \frac{\sqrt{1-y^2}}{y}\right)
\label{e3}
\end{eqnarray}
and the discrete states
\begin{equation}
\Delta E_{\rm discr} = -N(1-\nu) m \sqrt{1-y^2} 
\label{e4}
\end{equation}
can be taken over literally. Only the double counting correction of the interaction energy 
characteristic for the Hartree-Fock approach gets modified due to the bare quark mass $m_0$,
\begin{eqnarray}
\Delta E_{\rm d.c.} & = &  \int {\rm d}x \frac{(S-m_0)^2-(m-m_0)^2}{2g^2}
\nonumber \\
& = & \frac{1}{2g^2} \int {\rm d}x \left\{ (S^2-m^2)-2m_0 (S-m) \right\}.
\label{e5}
\end{eqnarray}
Moreover, when eliminating the bare coupling constant, we must now use the finite $m_0$ gap equation,
\begin{eqnarray}
\frac{1}{Ng^2} &=& \frac{1}{\pi}  \left(1- \frac{m_0}{m} \right)^{-1}    \ln  \frac{\Lambda}{m} 
\nonumber \\
& \approx & \frac{1}{\pi}  \left(1+ \frac{m_0}{m} \right) \ln  \frac{\Lambda}{m}  ,
\label{e6}
\end{eqnarray}
where we have dropped $m_0$-terms not matched by $\ln (\Lambda/m)$.
This reproduces the previous result plus a finite correction,
\begin{eqnarray}
\Delta E_{\rm d.c.} & = &  \frac{N}{2\pi} \ln \frac{\Lambda}{m}  \int {\rm d}x ( S^2-m^2) 
\nonumber \\
& & +  \frac{N}{2\pi} \frac{m_0}{m} \ln  \frac{\Lambda}{m}   \int {\rm d}x (S-m)^2 .
\label{e7}
\end{eqnarray}
Carrying out the integrations and introducing the physical parameter \cite{11}
\begin{equation}
\gamma = \frac{m_0}{\pi m} \ln \frac{\Lambda}{m}
\label{e8a}
\end{equation}
then yields
\begin{equation}
\Delta E_{\rm d.c.} = -  \frac{2Nym}{\pi} \ln \frac{\Lambda}{m} - 2 Nm \gamma ( y - {\rm arctanh}\, y) .
\label{e9}
\end{equation}
Collecting the results (\ref{e3}), (\ref{e4}) and (\ref{e9}), we get the following answer for the (variational) baryon mass
\begin{eqnarray}
\frac{M_B}{N} & = &  \frac{2ym}{\pi} + \frac{2m}{\pi} \sqrt{1-y^2} \arctan \frac{\sqrt{1-y^2}}{y}
\label{e10}  \\
& & -(1-\nu) m \sqrt{1-y^2} - 2m \gamma (y-{\rm arctanh}\, y).
\nonumber
\end{eqnarray}
We now choose $\nu$ (i.e., the fermion number) and vary $M_B$ with respect to $y$
\begin{equation}
\frac{\partial M_B}{\partial y }  = 0
\label{e11}
\end{equation}
or (discarding the trivial solution $y=0$)
\begin{equation}
\sqrt{1-y^2} \left( \frac{1}{\pi} \arctan \frac{\sqrt{1-y^2}}{y} - \frac{1-\nu}{2}\right) = \gamma y.
\label{e12}
\end{equation}
Introducing the angle $\theta$ via $y=\sin \theta$ ($0 \leq \theta \leq \pi/2$), we obtain
\begin{equation}
\frac{\nu}{2} = \frac{\theta}{\pi} + \gamma \tan \theta  .
\label{e13}
\end{equation}
If we eliminate $\nu$ from Eq.~(\ref{e10}) with the help of Eq.~(\ref{e13}),
the baryon mass at the minimum finally becomes
\begin{equation}
\frac{M_B}{N} = \frac{2m}{\pi} \sin \theta + 2 m \gamma \,{\rm arctanh}  (\sin \theta).
\label{e13a}
\end{equation}
The last two equations agree with Ref. \cite{11}. 

Now consider the self-consistency condition for the condensate and scalar potential
in the form
\begin{equation}
- \frac{(S-m_0)}{Ng^2} = \sum^{\rm occ} \bar{\psi} \psi .
\label{e14}
\end{equation}
The r.h.s. of Eq.~(\ref{e14}) gets contributions from the discrete states 
\begin{eqnarray}
\sum_{\rm discr}^{\rm occ} \bar{\psi} \psi &=& (1-\nu) \frac{m}{2} \sqrt{1-y^2} \left( \tanh \xi_- - \tanh \xi_+ \right)
\nonumber \\
& = & (1-\nu) \frac{\sqrt{1-y^2}}{2y} (S-m)
\label{e15}
\end{eqnarray}
and from the negative energy continuum,
\begin{eqnarray}
\sum_{\rm cont}^{\rm occ} \bar{\psi} \psi &=& - S \int_{-\Lambda/2}^{\Lambda/2}\frac{{\rm d}k}{2\pi}\frac{1}{E(k)}
 - m^3 y(1-y^2) 
\label{e16} \\
& & \cdot (\tanh \xi_- - \tanh \xi_+ )\int \frac{{\rm d}k}{2 \pi} \frac{1}{E(k)(k^2+m^2y^2)}
\nonumber \\
& = & - \frac{S}{\pi}  \ln \frac{\Lambda}{m} - \frac{(S-m)}{\pi}  \frac{\sqrt{1-y^2}}{y} \arctan \frac{\sqrt{1-y^2}}{y} ,
\nonumber
\end{eqnarray}
where we have again taken over results from Ref. \cite{14}.
The l.h.s. of Eq.~(\ref{e14}) can be rewritten with the help of the gap equation (\ref{e6}),
\begin{eqnarray}
- \frac{(S-m_0)}{Ng^2} &=& -  \frac{S}{\pi} \ln \frac{\Lambda}{m} -  \frac{S-m}{\pi} \frac{m_0}{m} \ln \frac{\Lambda}{m}
\nonumber \\
& = & - \frac{S}{\pi} \ln \frac{\Lambda}{m} - (S-m) \gamma .
\label{e17}
\end{eqnarray}
Combining Eqs.~(\ref{e15}), (\ref{e16}) and (\ref{e17}), the self-consistency condition assumes the form
\begin{equation}
\frac{1- \nu}{2} \frac{\sqrt{1-y^2}}{y} - \frac{1}{\pi} \frac{\sqrt{1-y^2}}{y} \arctan \frac{\sqrt{1-y^2}}{y}
+ \gamma  = 0
\label{e18}
\end{equation}
which coincides with the variational equation (\ref{e12}). Therefore, surprisingly, the variational potential turns out to be
self-consistent and hence to provide us with the exact baryon of the massive GN model in the large $N$ limit.
We have been rather explicit here because although our variational calculation agrees with Ref. \cite{11},
we have come to a different conclusion concerning the self-consistency of this ansatz. 

Let us summarize our findings. In the massless limit, the shape of the self-consistent scalar potential 
depends only on the filling of the valence level. For complete filling ($\nu=1$), the well degenerates into
infinitely far separated kink and antikink. As one decreases $\nu$, the baryon shrinks and eventually
approaches a $1/\cosh^2$-shape as expected from the non-linear Schr\"odinger equation \cite{14}.
In the massive Gross-Neveu model, we have another parameter $\gamma$ which controls the shape of the
scalar potential. If we consider only complete filling and turn on the bare mass term respectively $\gamma$,
the baryon also shrinks and eventually becomes non-relativistic. This is physically very reasonable. Interestingly, no
new shape of $S(x)$ appears as we turn on the bare quark mass. It is this feature which makes
the massive GN model exactly solvable, a fact which we had not expected.
The only question is how the parameter $\theta$ in the general kink-antikink potential depends
on $\nu$ and $\gamma$. This is answered (implicitly) by Eq.~(\ref{e13}). Eq.~(\ref{e13a}) then yields the
baryon mass as a function of fermion number $\nu$ and symmetry breaking parameter $\gamma$. 
%################################################################################################
%                                                                                           SECTION 3
%################################################################################################
\section{Baryons in the massive NJL$_2$ model --- derivative expansion}
In the present section, we consider the NJL$_2$ model with continuous chiral symmetry. Unlike in the
case of the GN model, here we have not been able to guess the self-consistent
Hartree-Fock potential for arbitrary bare quark masses.
The only clue we have is the fact that near the chiral limit,
the sine-Gordon model should somehow emerge. In this limit, we know that the relevant length scale over
which the potential varies is given by $1/\mu$, the inverse pion mass.
This suggests that the theoretical instrument of choice should be the derivative expansion, at
least in the vicinity of the chiral limit. We will see that the sine-Gordon equation is nothing but
the leading order term of such a derivative expansion. Moreover, the calculation of higher order corrections can
be done in closed analytical form, so that we will get a quantitative understanding of the baryons in a reasonable
range of bare mass parameters, although not for arbitrary mass. This is a nice example where one can study
in all detail how a Skyrme-type description of baryons emerges from an underlying fermionic theory, including
an unusually good control of the corrections.

As far as the derivative expansion is concerned, we refer the reader to the original paper \cite{15}
as well as to Ref. \cite{16} from which we take over the basic formulae and the notation.
Nevertheless, we shall try to keep our paper self-contained as far as possible.  

To avoid confusion, we should like to draw the attention to the following difference between this section
and the preceding one. In the case of the NJL$_2$ model, the Skyrme-picture applies only to the case
where the fermion number is an integer multiple of $N$, the number of flavours (this correponds to
$\nu=1$ in Sect. II). One should not think of filling a positive energy valence level, but rather 
that the winding number of the potential induces fermion number by sending energy levels from
positive to negative energies or vice versa, changing the spectral asymmetry. This is discussed
in more detail in Refs. \cite{2,10}. Only for this case of complete filling of single particle states, the baryon gets massless
in the chiral limit. One can also contemplate partially filled states in the NJL$_2$ model, along the lines of the 
original work of Shei \cite{17}, but we shall not consider this possibility here. 
This allows us in the following to define the baryon number as an integer, namely fermion number
divided by $N$.

\subsection{General setup of the derivative expansion}
The central quantity for computing ground state energy and baryon number for a system with Hamiltonian $H$ 
is the spectral density
\begin{equation}
\sigma(E) = {\rm Tr}\,  \delta(H-E) = \frac{1}{\pi} {\rm Im} \,{\rm Tr} R(E+{\rm i}\epsilon),
\label{d1}
\end{equation}
where we have introduced the resolvent 
\begin{equation}
R(z) = {\rm Tr}\, \frac{1}{H-z} = {\rm Tr}\, \frac{H+z}{H^2-z^2}.
\label{d2}
\end{equation}
The induced fermion number is ($-1/2$) times the spectral asymmetry,
\begin{eqnarray}
\langle N \rangle &=&  - \frac{1}{2} \int_{- \infty}^{\infty} {\rm d}E\, \sigma(E) \, {\rm sgn}(E)
\label{d3} \\ 
& = & - \frac{1}{2\pi} {\rm Im} \int_0^{\infty} {\rm d}E \left[ R(E+{\rm i}\epsilon )+ R(-E-{\rm i}\epsilon )\right],
\nonumber
\end{eqnarray}
whereas the ground state energy can be written as 
\begin{eqnarray}
\langle H \rangle & = & \int_{- \infty}^0{\rm d}E\, E\, \sigma(E)
\nonumber \\ 
& = &  \frac{1}{\pi} {\rm Im} \int_{- \infty}^0 {\rm d}E\, E\, R(E+{\rm i} \epsilon)
\label{d4}
\end{eqnarray}
In a notation similar to Ref. \cite{16} we decompose $H$ and $H^2$ formally into
\begin{equation}
H = K + I \ , \qquad H^2 = H_0^2 + V
\label{d5}
\end{equation}
and expand the resolvent in powers of $V$,
\begin{eqnarray}
R(z) & = & {\rm Tr} (K+I+z)\frac{1}{H_0^2-z^2+V}
\nonumber \\
& = & {\rm Tr} (K+I+z)\left( G_0 \sum_{n=0}^{\infty}(-VG_0)^n\right)
\label{d6}
\end{eqnarray}
with
\begin{equation}
G_0 = \frac{1}{H_0^2-z^2} .
\label{d7}
\end{equation}
If one commutes the $V$'s through the $G_0$'s by repeatedly applying the identity
\begin{equation}
G_0 V = V G_0 + G_0 \left[ V,H_0^2 \right] G_0,
\label{d8}
\end{equation}
one generates the derivative expansion (the commutator $[V,H_0^2]$ involves derivatives
of $V$). As is well known, this method quickly becomes tedious due to the proliferation of higher order terms.
Therefore we shall use another technique based on momentum space below.

\subsection{Specialization to the massive NJL$_2$ model} 
In the large $N$ limit, the Hartree-Fock approach is adequate. We can therefore choose for $H$ a 
single particle Dirac Hamiltonian with general scalar and pseudoscalar local potentials.
Without loss of generality, it can be cast into the convenient form 
\begin{eqnarray}
H &=& -  \gamma^5 {\rm i} \partial_x + m_0 \gamma^0
\nonumber \\
& & + \left[\bar{m} + \lambda(x) \right]{\rm e}^{{\rm i}\gamma^5 \chi(x)}\gamma^0
 {\rm e}^{-{\rm i}\gamma^5 \chi(x)}
\label{d9}
\end{eqnarray}
which exhibits the bare mass term and a Hartree-Fock potential equivalent to
\begin{equation}
\sigma(x) \gamma^0 + \pi(x) {\rm i} \gamma^1
\label{d10}
\end{equation}
with
\begin{eqnarray}
\sigma & = & (\bar{m} + \lambda)\cos 2 \chi 
\nonumber \\
\pi & = & - (\bar{m}+ \lambda)\sin 2 \chi.
\label{d11}
\end{eqnarray}
We have introduced a mass parameter $\bar{m}$ which differs from the physical fermion mass $m$,
\begin{equation}
m = \bar{m} + m_0
\label{d14}
\end{equation}
to ensure that $\lambda(x)$ vanishes asymptotically.
The $\gamma$-matrices will be taken in the representation $\gamma^0 = \sigma_1, \gamma^1 = -{\rm i} \sigma_2,
\gamma^5=\sigma_3$. Next we identify the operators $K$ and $I$ from the preceding subsection as follows,
\begin{equation}
H = K + I = \left( \begin{array}{cc} -{\rm i} \partial_x & 0 \\ 0 & {\rm i} \partial_x \end{array} \right)
+ \left( \begin{array}{cc}0 & \Phi \\ \Phi^* & 0 \end{array} \right)
\label{d12}
\end{equation}
with 
\begin{equation}
\Phi = (\bar{m} + \lambda ) {\rm e}^{2{\rm i}\chi} + m_0.
\label{d13}
\end{equation}
A natural way of decomposing $H^2$ into $H_0^2$ and $V$ is
\begin{equation}
H_0^2 = \left( - \partial_x^2 + m^2 \right) 
\label{d15}
\end{equation}
\begin{equation}
V = \left( \begin{array}{cc}|\Phi|^2 - m^2  & - {\rm i}\Phi' \\ {\rm i}(\Phi')^* & |\Phi|^2-m^2  \end{array} \right)
\label{d16}
\end{equation}
where, using Eq.~(\ref{d13}),
\begin{eqnarray}
|\Phi|^2-m^2 & = & 2 \bar{m} \lambda + \lambda^2 + 2m_0(\bar{m}+\lambda) \cos 2 \chi - 2 m_0 \bar{m}
\nonumber \\
\Phi' & = & \left( \lambda' + 2 {\rm i} (\bar{m}+\lambda ) \chi' \right) {\rm e}^{2 {\rm i} \chi}.
\label{d17}
\end{eqnarray}
We shall perform the derivative expansion of the basic building blocks
\begin{equation}
{\rm Tr} (K+I+z) G_0 (VG_0)^n
\label{d18}
\end{equation}
of the resolvent as follows. Consider first the term proportional to $z$. 
In momentum space,
\begin{eqnarray}
{\rm Tr}  G_0 (VG_0)^n &=& \int \frac{{\rm d}p}{2\pi} \frac{{\rm d}q_1}{2\pi} ... \frac{{\rm d}q_{n-1}}{2\pi}
G_0(p)^2G_0(p+q_1)... 
\nonumber \\
G_0(p+q_{n-1}) &{\rm tr}& V(q_1) V(q_2-q_1)...V(-q_{n-1})
\label{d19}
\end{eqnarray}
where now tr is only the Dirac trace. In $p$-space, the potentials vary rapidly as compared to the Green's functions.
We can therefore expand the product of $G_0$'s in a power series in the $q_i$'s. Once this is done, we
transform the potentials back to coordinate space and carry out most of the integrations. The $q_i$'s
are replaced by derivatives acting on the $V's$,
 \begin{eqnarray}
{\rm Tr}  G_0 (VG_0)^n &=&   \int {\rm d}x \int \frac{{\rm d}p}{2\pi} G_0(p)^2G_0(p+q_1)...
\nonumber \\
& & \left.G_0(p+q_{n-1})  \right|_{q_k=i(\partial_1+...+\partial_{k})}
\nonumber \\
& & \cdot \,{\rm tr} \left. V(x_1)V(x_2)...V(x_{n})\right|_{x_k=x}.
\label{d20}
\end{eqnarray}
In this process, partial integrations are carried out which can be justified for the functions $\lambda,\chi$ 
which will come out at the end (there are no surface terms).
We have suppressed the Taylor expansion in Eq.~(\ref{d20}) in order to keep the structure of the formula transparent. 
The other two terms in Eq.~(\ref{d18}) can be handled similarly with the result
\begin{eqnarray}
{\rm Tr} I G_0 (VG_0)^n &=&   \int {\rm d}x \int \frac{{\rm d}p}{2\pi} G_0(p) G_0(p+q_1)...
\nonumber \\
& & \left. G_0(p+q_{n}) \right|_{q_k=i(\partial_1+...+\partial_{k})}
\nonumber \\
& & \cdot \,{\rm tr} \left. I(x_1) V(x_2)V(x_3)...V(x_{n+1})\right|_{x_k=x}
\nonumber \\
{\rm Tr} K G_0 (VG_0)^n &=&   \int {\rm d}x \int \frac{{\rm d}p}{2\pi} p\, G_0(p)^2G_0(p+q_1)...
\nonumber \\
& & \left. G_0(p+q_{n-1}) \right|_{q_k=i(\partial_1+...+\partial_{k})}
\nonumber \\
& & \cdot \,{\rm tr} \left. \sigma_3 V(x_1)V(x_2)...V(x_{n})\right|_{x_k=x}.
\label{d21}
\end{eqnarray}
If we apply the above formalism to Hartree-Fock, we must of course add the double counting correction to the energy,
\begin{equation}
E = \langle H \rangle +  \int {\rm d}x \frac{(\bar{m}+ \lambda)^2}{2Ng^2}.
\label{d22}
\end{equation}

\subsection{Energy}
Only the odd part of the resolvent $R(z)$ in Eq.~(\ref{d6}) contributes. In the expansion in powers of $V$,
the terms of order $V^0$ and $V^1$ play a special role. They contain divergencies which require 
regularization and renormalization, whereas all higher order terms are finite. Besides, they do not
necessitate the derivative expansion (there are no $q_i$ to expand in). We therefore show this calculation
in some detail first and then discuss the higher order terms.

{\em Terms of order $V^0$}

Here we are considering the vacuum energy. Although this will be subtracted from the baryon energy,
it is useful to calculate it in order to rederive the renormalization condition needed later on. The resolvent
is given by 
\begin{equation}
R(z)=\int {\rm d}x \int_{-\Lambda/2}^{\Lambda/2} \frac{{\rm d}p}{2\pi} \frac{2z}{(p^2+m^2)-z^2}
\label{d23}
\end{equation} 
with an UV cut-off $\Lambda/2$. Eqs.~(\ref{d4}) and (\ref{d22}) then yield the energy density
\begin{eqnarray}
{\cal E}_{\rm vac} &=& - \int_{-\Lambda/2}^{\Lambda/2} \frac{{\rm d}p}{2\pi} \sqrt{p^2+m^2} + \frac{\bar{m}^2}{2Ng^2}
\nonumber \\
& = & - \frac{\Lambda^2}{8\pi} - \frac{m^2}{4\pi} + \frac{m^2}{2\pi} \ln \frac{m}{\Lambda} + \frac{(m-m_0)^2}{2 Ng^2}.
\label{d24}
\end{eqnarray}
Minimizing with respect to the dynamical fermion mass $m$, we recover the well-known gap equation
\begin{equation}
\frac{(m-m_0)}{Ng^2} = \frac{m}{\pi} \ln \frac{\Lambda}{m}.
\label{d25}
\end{equation}
Upon using this condition, the vacuum energy density becomes (up to an irrelevant quadratic divergence)
\begin{equation}
{\cal E}_{\rm vac} = - \frac{m^2}{4\pi} - \frac{m m_0}{2\pi} \ln \frac{\Lambda}{m}.
\label{d26}
\end{equation}
As pointed out above, $m_0/Ng^2$ or equivalently $m_0 \ln \Lambda/m$ is a physical quantity.
Here, it is more natural to express it in terms of the pion mass $\mu$ (see \cite{7}),
\begin{equation}
\frac{\pi m_0}{Ng^2 m} = \frac{1}{\sqrt{\eta-1}} \arctan  \frac{1}{\sqrt{\eta -1}} \ , \quad \eta = \frac{4m^2}{\mu^2}.
\label{d27}
\end{equation}
For later convenience, we introduce the function
\begin{eqnarray}
F(y) &=& \frac{4}{y\sqrt{4-y^2}} \arctan  \frac{y}{\sqrt{4-y^2}} 
\nonumber \\
& = & 1+ \frac{1}{6} y^2+ \frac{1}{30}y^4 + \frac{1}{140} y^6 + ...
\label{d28}
\end{eqnarray}
which allows us to express the renormalized vacuum energy density in the following way,
\begin{equation}
{\cal E}_{\rm vac} = - \frac{m^2}{4\pi} - \frac{\mu^2}{8\pi} F(\mu/m).
\label{d29}
\end{equation}
This already shows that an expansion for small quark masses should be thought of as an expansion in 
the ratio of pion mass to (dynamical) quark mass in the NJL$_2$ model. The bare quark mass $m_0$ goes to 0 in the
limit $\Lambda \to \infty$ and cannot appear in any physical quantity.

{\em Terms of order $V^1$}

Now we have to compute
\begin{eqnarray}
\langle H \rangle^{(1)} &=& - \frac{1}{\pi} \int_{-\infty}^0 {\rm d}E\,  E\, {\rm Im}\, {\rm Tr}\, z G_0VG_0 
\nonumber \\
& & +\int {\rm d}x  \frac{(\bar{m}+\lambda)^2 -\bar{m}^2}{2Ng^2}
\nonumber \\
& = &  \int {\rm d}x \left\{ - \frac{1}{\pi}\int_{-\Lambda/2}^{\Lambda/2} \frac{{\rm d}p}{2\pi} \int_{-\infty}^0 {\rm d}E\, E\,
\right.
\label{d30} \\
& & \cdot \,{\rm Im}\, z G_0^2(p)\, {\rm tr} \,V(x) 
+ \left. \frac{(\bar{m}+\lambda)^2 -\bar{m}^2}{2Ng^2}\right\}.
\nonumber
\end{eqnarray}
The integrations over $E$ and $p$ can easily be carried out. Owing to the gap equation,
all divergencies cancel out and the unphysical quantities
($Ng^2, m_0, \Lambda$) can again be eliminated in favour of $\mu$ and $m$ with the result
\begin{eqnarray}
\langle H \rangle^{(1)} &=&  \frac{\mu^2}{4\pi} F(\mu/m)
\label{d31} \\
& &  \cdot \int {\rm d}x\left[\left(1+\frac{\lambda}{m} \right) \left(1-\cos 2\chi \right)+
\frac{\lambda^2}{2m^2}\right] .
\nonumber
\end{eqnarray}
Once again we have dropped all terms in which the bare mass $m_0$ is not multiplied
by a factor $\ln (\Lambda/m)$.

{\em Terms of order $V^n$, $n>1$}

All higher order terms in the expansion (\ref{d6}) are free of UV divergencies. Therefore, we can set the bare
quark mass $m_0=0$, thereby greatly simplifying our potential $V$, and calculate the traces
and integrals mechanically with computer algebra (we used Maple). The result for the energy density coming
from these higher order terms is a polynomial in $\lambda$, $\chi$ and their derivatives. Detailed results will be given in 
Sect. III E.

\subsection{Baryon number}
The leading order contribution comes from the ${\rm Tr} IG_0VG_0$ term in Eq. ({\ref{d6}) and can easily be shown to have the
expected topological form
\begin{equation}
\langle N \rangle = \int {{\rm d}x} \frac{\chi'}{\pi} = \frac{1}{\pi} \left[ \chi(\infty) - \chi(-\infty)\right],
\label{d32}
\end{equation}
with the well-known identification of winding number of the chiral phase with baryon number.
Since we were not sure to which extent the general
topological arguments carry over to a system with explicit symmetry breaking, we have computed the next three
orders in the derivative expansion, using Maple. We found that the result (\ref{d32}) does not get 
any correction whatsoever and therefore presumably holds to all orders. Although many terms are produced
in the integrand (the baryon density) by our algorithm, they can be nicely combined into total derivatives of functions
which vanish at infinity and therefore do not affect the lowest order topological result (\ref{d32}).

\subsection{Results}
On the basis of the effective sine-Gordon theory \cite{9} for the NJL$_2$ model, the mass of the baryon
with fermion number $N$  is expected to approach $2\mu/\pi$ for $\mu \to 0$. Our goal is to compute corrections
to this result up to order $\mu^6/m^6$. This requires a substantial number of terms in the derivative expansion
and correspondingly involved algebraic expressions. To simplify the notation, let us set $m=1$ from now on.
Then, the energy density in the relevant order of the derivative expansion computed along the lines of 
Sect. III C has the following form,
\begin{eqnarray}
2 \pi {\cal E} &=&  -\frac{\mu^2}{2} F(\mu)
\left[ \left(1 + \lambda \right) \left(\cos 2 \chi -1 \right)- \frac{1}{2}\lambda^2 \right] 
\nonumber \\
& & +  (\chi')^2 - \frac{1}{6} (\chi'')^2 + \frac{1}{30}(\chi''')^2- \frac{1}{140}(\chi^{IV})^2 
\nonumber \\
& &  - \frac{1}{45} (\chi'')^4 +  \lambda^2 + \frac{1}{12} (\lambda')^2 - \frac{1}{120}(\lambda'')^2
\nonumber \\
& &  + \frac{1}{3} \lambda^3 - \frac{1}{6} \lambda (\lambda')^2 - \frac{1}{12} \lambda^4
 + \frac{1}{3} \lambda (\chi'')^2
\nonumber \\
& &   + \frac{1}{15} \lambda (\chi''')^2 + \frac{1}{5}\lambda \chi'' \chi^{IV}
  - \frac{1}{2} \lambda^2 (\chi'')^2
\label{d34}
\end{eqnarray}
We have subtracted the vacuum contribution and simplified the result as much as possible with
the help of partial integrations (only the energy $\int {\rm d}x\,  {\cal E}$ is uniquely
defined in this approach). To appreciate the complexity of expression (\ref{d34}), 
one should compare it with the leading order terms only,
\begin{equation}
2 \pi {\cal E}  =  (\chi')^2 - \frac{\mu^2}{2} (\cos 2 \chi -1)  ,
\label{d35a}
\end{equation}
which reproduce exactly the sine-Gordon theory and represent the state of the art
prior to this work. We then vary
the energy with respect to $\chi$ and $\lambda$.
In order to solve the resulting differential equations, we expand $F(\mu)$ and assume the following
Taylor series for $\chi$ and $\lambda$, 
\begin{eqnarray}
\chi & \approx & \chi_0 +  \mu^2 \chi_1 + \mu^4 \chi_2 + \mu^6 \chi_3 
\nonumber \\
\lambda & \approx & \mu^2 \lambda_1 +  \mu^4 \lambda_2 +  \mu^6 \lambda_3.
\label{d35}
\end{eqnarray}
The coefficients $\chi_n$ and $\lambda_n$ in turn are taken to depend only on $\xi=\mu x$, so that each derivative
with respect to $x$ increases the power of $\mu$ by one.  All of these assumptions can be justified {\em a posteriori}
by showing that they are the simplest ones which lead to a consistent approximate solution of the differential equations.
This procedure yields the following set of inhomogeneous differential equations for $\chi_n$ and algebraic equations
for $\lambda_n$ (where now $\,' = {\rm d}/{\rm d}\xi$),
\begin{widetext}
\begin{eqnarray}
\chi_0'' & = & \frac{1}{2} \sin 2 \chi_0
\label{d36} \\
\lambda_1 & = & \frac{1}{4} \left( \cos 2 \chi_0 - 1 \right)
\label{d37} \\
\chi_1'' -   \chi_1 \cos 2 \chi_0   &=& \frac{1}{2}\left( \lambda_1+ \frac{1}{6} \right) \sin 2 \chi_0
- \frac{1}{6}\chi_0^{IV}
\label{d38} \\
\lambda_2 & = & \frac{1}{24} \left( \cos 2 \chi_0 -1 \right) - \frac{1}{2} \chi_1 \sin 2 \chi_0
\nonumber \\
& & - \frac{1}{6}(\chi_0'')^2 - \frac{1}{4} \lambda_1 - \frac{1}{2}\lambda_1^2 +\frac{1}{12}\lambda_1''
\label{d39} \\
\chi_2'' -  \chi_2  \cos 2 \chi_0  & = & \left( - \chi_1^2+ \frac{1}{2}\lambda_2+ \frac{1}{12} \lambda_1  + 
\frac{1}{60}\right) \sin 2 \chi_0
\nonumber \\
& & +  \left( \lambda_1 \chi_1 + \frac{1}{6} \chi_1 \right) \cos 2 \chi_0- \frac{1}{30}\chi_0^{VI}
\nonumber \\
& & + \frac{1}{3} \lambda_1 \chi_0^{IV} + \frac{1}{3} \lambda_1'' \chi_0''+\frac{2}{3} \lambda_1' \chi_0''' 
- \frac{1}{6}\chi_1^{IV}
\label{d40} \\
\lambda_3 & = &  - \left( \frac{1}{2} \chi_1^2 - \frac{1}{120}\right) \cos 2 \chi_0
-\left( \frac{1}{2} \chi_2 + \frac{1}{12} \chi_1\right) \sin 2 \chi_0
\nonumber \\
& & - \lambda_1 \lambda_2 - \frac{1}{120} - \frac{1}{10}\chi_0'' \chi_0^{IV}
- \frac{1}{30} (\chi_0''')^2-\frac{1}{4} \lambda_2 + \frac{1}{12}\lambda_2''+ \frac{1}{6} \lambda_1^3
\nonumber \\
& & - \frac{1}{3}\chi_0'' \chi_1'' - \frac{1}{12} (\lambda_1')^2
- \frac{1}{6}\lambda_1 \lambda_1'' + \frac{1}{120} \lambda_1^{IV}
- \frac{1}{24} \lambda_1 + \frac{1}{2}\lambda_1 (\chi_0'')^2
\label{d41} \\
\chi_3''-    \chi_3  \cos 2 \chi_0  & = &  \left( \frac{1}{6} \chi_2 + \lambda_1 \chi_2+ 
 \lambda_2  \chi_1- \frac{2}{3}  \chi_1^3+ \frac{1}{30} \chi_1+ \frac{1}{6} \lambda_1 \chi_1 
\right) \cos 2 \chi_0
\nonumber \\
& & -   \left( 2  \chi_1 \chi_2- \frac{1}{280} + \frac{1}{6} \chi_1^2  - \frac{1}{12} \lambda_2 
- \frac{1}{2} \lambda_3  +  \lambda_1 \chi_1^2 - \frac{1}{60} \lambda_1 \right) \sin 2 \chi_0
\nonumber \\
& & -   \frac{1}{140} \chi_0^{VIII}- \frac{4}{15} \chi_0'' (\chi_0''')^2- \frac{2}{15}  (\chi_0'')^2 \chi_0^{IV} 
+ \frac{1}{2}\lambda_1'' \chi_0^{IV}+ \frac{1}{3} \lambda_1''' \chi_0'''
\nonumber \\
& & +   \frac{2}{15}  \lambda_1 \chi_0^{VI}
+ \frac{2}{5} \lambda_1' \chi_0^{V}+\frac{1}{10} \lambda_1^{IV}\chi_0''- \frac{1}{6} \chi_2^{IV}
+ \frac{1}{3} \lambda_2'' \chi_0'' +  \frac{2}{3}  \lambda_1' \chi_1'''
\nonumber \\
& & +    \frac{2}{3}  \lambda_1' \chi_1'''+ \frac{1}{3} \lambda_1 \chi_1^{IV}
+ \frac{2}{3} \lambda_2' \chi_0'''
+ \frac{1}{3} \lambda_1 \chi_1^{IV}
+ \frac{2}{3} \lambda_2' \chi_0'''
+ \frac{1}{3}\lambda_2 \chi_0^{IV} 
\nonumber \\
& & +   \frac{1}{3} \lambda_1'' \chi_1''
-\frac{1}{30} \chi_1^{VI}- \lambda_1  \lambda_1'' \chi_0''- 2 \lambda_1  \lambda_1' \chi_0'''
 - \frac{1}{2}\lambda_1^2 \chi_0^{IV} -  (\lambda_1')^2 \chi_0''  
\label{d42}
\end{eqnarray}
\end{widetext}
In spite of their frightening appearance, it is not too hard to solve these equations recursively
in the order in which they have been written down here.
Our analytical results for baryon number 1 are (${\rm sech}\, \xi = 1/\cosh \xi$)
\begin{eqnarray}
\chi_0 &=& 2 \arctan {\rm e}^{\xi}
\nonumber \\
\lambda_1 & = & -  \frac{1}{2} {\rm sech}^2 \xi 
\nonumber \\
\chi_1 & = & \frac{1}{8} \sinh \xi  \,{\rm sech}^2 \xi 
\nonumber \\
\lambda_2 & = & - \frac{1}{6} \left( {\rm sech}^2 \xi  - {\rm sech}^4 \xi \right)
\nonumber \\
\chi_2 & = & - \frac{1}{1152} \sinh \xi \left( 23 \,{\rm sech}^2 \xi  - 122 \,{\rm sech}^4 \xi  \right)
\nonumber \\
\lambda_3 & = & - \frac{1}{1440} \left( 252 \,{\rm sech}^2 \xi  - 1225 \,{\rm sech}^4 \xi  + 1061 \,{\rm sech}^6 \xi \right) 
\nonumber
\end{eqnarray}
\begin{eqnarray}
\chi_3 & = & - \frac{1}{691200} \sinh \xi \left( 2621 \,{\rm sech}^2 \xi  - 108092 \,{\rm sech}^4 \xi \right.
\nonumber \\
& & \left. + 90456 \,{\rm sech}^6 \xi  \right)
\label{d43}
\end{eqnarray}
In the course of this calculation, we had to decide what to do with the homogeneous solution
of the differential equations for $\chi_n$. Physically, they reflect the fact that there is a flat direction
in function space due to the breaking of translational invariance by the baryon. We have made the choice that 
$\chi$ is odd under $\xi \to - \xi $, a requirement which fixes the position of the baryon in space and leads to 
a unique solution of the differential equations.

We see that the leading order results (keeping only $\chi_0$) agree with the sine-Gordon theory, whereas the 
higher order corrections yield systematic corrections to it. Inserting Eqs.~(\ref{d43}) into
Eqs.~(\ref{d35}), the scalar and pseudoscalar potentials can be written as a power series in $\mu^2$,
i.e., the ratio of pion mass to physical quark mass (remember that we have chosen units such that $m=1$),
with smooth coefficient functions depending only on $\mu x$.

Inserting the results (\ref{d43}) into the expression for the energy density (\ref{d34}) and integrating over ${\rm d}x$, we finally
get the following chiral expansion for the baryon mass in the NJL$_2$ model,
\begin{equation}
\frac{M_B}{N} = \frac{2 \mu}{\pi} \left\{ 1 - \frac{1}{36} \mu^2 - \frac{1}{300} \mu^4 - \frac{1}{588} \mu^6 + {\rm O}(\mu^8) \right\}
\label{d44}
\end{equation}
Due to the values of the coefficients, the corrections are numerically small even for $\mu/m \sim 1$. 
We see no difficulty in principle to push this calculation to higher orders, but the number of terms appearing
in intermediate steps of the calculation increases rapidly. It then becomes increasingly difficult for Maple 
to handle and simplify the lengthy expressions.
%################################################################################################
%                                                                                           SECTION 4
%################################################################################################
\section{Summary and conclusions}
The purpose of this work was to begin a systematic study of massive Gross-Neveu models in 1+1 dimensions.
As a warm-up for a comprehensive study of the phase diagram of
these models, we have reconsidered the issue of baryons. Two different strategies turned out to
be successful for the massive discrete chiral GN model and the massive continuous chiral NJL$_2$ model,
respectively.

In the case of the GN model, we simply guessed the scalar potential for a Hartree-Fock calculation
and proved its self-consistency. Actually, the shape of the potential is the same as in the massless limit, but
for a different fermion number. This was crucial for being able to carry through the calculation, since 
potentials for which one can solve the Dirac equation analytically are extremely rare. The same
calculation in a somewhat different framework has already been done some time ago \cite{11}; however, the
decisive fact that the result is self-consistent was apparently missed by the authors. This baryon solution should be a good
starting point for studying hot and dense matter in the GN model, generalizing our recent work 
in the chiral limit \cite{6} to finite bare quark masses.

Baryons in the NJL$_2$ model are a totally different story: In the chiral limit, they become massless, 
baryon number has a topological meaning, and a Skyrme-type physical picture is adequate. Here we
applied a derivative expansion technique which allows us to bypass the cumbersome Hartree-Fock procedure.
Without explicitly solving the Dirac equation, we have nevertheless obtained quantitative results 
encoded in the first three correction terms beyond the sine-Gordon limit. In effect, this procedure can be regarded 
as a kind of bosonization where a chiral angle field $\chi$ and a radial field $\lambda$ carry all the dynamical
information. The emergence of the Skyrme picture is put on very solid grounds in this case. The small parameter which
governs the expansion of potentials and baryon masses is the ratio of pion mass to dynamical quark mass,
whereas the bare coupling constant and the bare quark mass disappear in the process of renormalization. 

Unfortunately, it is unlikely that this method carries over to higher dimensions. The fact that the size of the
baryon increases like the inverse pion mass in the NJL$_2$ model was instrumental here, but is not expected
to hold in more than one space dimension. Nevertheless, we hope that our results are of some use to test algorithms or
study questions related to chiral perturbation theory with baryons.

Finally, let us comment on the relationship between the GN model and quasi-one-dimensional condensed matter systems.
As is well known and has recently been re-emphasized by us \cite{6}, the GN model can serve to describe systems like
the Peierls-Fr\"ohlich model, one-dimensional superconductors, or conducting polymers. Does the step from the
massless to the massive GN model have any analogy as well? The answer is yes --- it is the
transition from systems with a two-fold degenerate ground state to non-degenerate systems. The prime example is
perhaps the step from trans-polyacetylene to cis-polyacetylene. Following the work
of Brazovskii and Kirova \cite{18}, a number of studies have been devoted to the existence of polarons, bipolarons and
excitons in such doped, non-degenerate polymers (doping changes the number of electrons in the valence band), 
see the reprint volume \cite{19}.
We can compare these studies to our results in Sect. II for $N=2$ (there is no flavour in the condensed matter case, 
only two spin states). Thus, the bipolaron corresponds to our $\nu=1$ case, the polaron to $\nu=1/2$, and
indeed for these values the mathematics of the condensed matter continuum models is identical to our calculation.
In the polymer case, one also considers states with one electron in the negative and positive energy valence levels,
an exciton. As a matter of  fact, our calculation can be repeated for arbitrary occupation fractions $\nu_{\pm}$
for these two discrete levels, preserving self-consistency. Such states would correspond to heavier, excited
baryons in the GN model, with fermion number $N(\nu_+ + \nu_- -1)$ and an admixture of anti-quarks.
If the baryon number vanishes ($\nu_+ = 1- \nu_-$), one could perhaps talk about ``baryonium" as the analogue of an
exciton.

\end{document}